\documentstyle[12pt]{article}
\begin{document}
\newcommand{\ot}{\frac{1}{2}}
\newcommand{\D}{& \displaystyle}  %instead of &
\newcommand{\di}{\displaystyle}   %at begin of each line
\font \math=msbm10 scaled \magstep 1
% This is not accessed in math mode: use it as e.g. \mbox{\math Z}}
\newcommand{\mmath}[1]{{\mbox{\math #1}}}
\renewcommand{\Re}{\mmath{R}}
\newcommand{\ka}{\kappa}
\newcommand{\be}{\beta}
\newcommand{\kac}{\kappa_c}
\newcommand{\psibar}{\overline{\psi}}
\newcommand{\myr}{\mbox{\math R}}
\newcommand{\myc}{\mbox{\math C}}

\date{August 1994}
\title
{Lee-Yang zeroes in the one flavour massive lattice Schwinger model}
\author
{\bf H. Gausterer and C.B. Lang \\  \\
Institut f\"ur Theoretische Physik,\\
Universit\"at Graz, A-8010 Graz, AUSTRIA}
\maketitle
\begin{abstract}

We study the partition function of the model formulated with Wilson
fermions with only one species, both analytically and numerically. At
strong coupling we construct the solution for lattice size up to
$8\times 8$, a polynomial in the hopping parameter up to
$O(\ka^{128})$.  At $\be>0$ we evaluate the expectation value of the
fermion determinant for complex values of $\ka$. From the Lee-Yang
zeroes we find support for the existence of a line of phase transitions
from $(\be=0, \ka\simeq 0.38)$ up to $(\be=\infty, \ka=1/4)$.

\end{abstract}
\newpage

\section{Introduction}

QED$_2$, the theory of electrons and photons in 2D, for massless
electrons is analytically solvable \cite{Sc62} and has been studied
extensively \cite{ScPaps}. In the original version the system has
$n_f=1$ fermions. The fermions are confined and the model is equivalent
to a system of non-interacting bosons.  The theory is
superrenormalizable and there is only finite renormalization of the
charge. Chiral symmetry is broken due to an anomaly in the axial
current thus there is no Goldstone boson, but a massive pseudoscalar of
Gaussian nature. One has charge shielding, i.e.  there are no
long-range forces between static charges.  Allowing for $n_f>1$
fermions, one does have additional massless states \cite{GaSe94}, a
situation prototyped in the $O(n_f)$ non-linear $\sigma$-model.

Many of these properties are intriguingly similar to those of QCD in
4D, which in the non-perturbative domain is studied mainly in  the
lattice regularization by computer simulations.  The Schwinger model
formulated on a lattice is therefore a challenging 2D model for
lattice QCD$_4$. In this formulation it naturally encompasses the
massless and the massive situation. In the continuum the massive
Schwinger model cannot be solved explicitly although there are results
perturbative in $m/e$ \cite{CoJaSu75,FrSe76,CaKe86}.  From these we
expect a situation similar to the massless case with further bound
states and quark trapping, again very much like in QCD$_4$.

Lattice formulations of fermions are plagued with the doubling
problem.  There are more than one pole in the Brillouin zone of the
momentum space propagator. In the staggered (Kogut-Susskind)
formulation one distributes
different components of the spinor on different sites of a hypercube
and thereby effectively reduces the $2^D$ multiplicity by a factor
$2^{[D/2]}$.  In the Wilson formulation the doubler modes are given
masses of $O(1/a)$ and one expects decoupling in the continuum limit
$a\to 0$.  Furthermore most lattice calculations rely on the hybrid
Monte Carlo method to implement the fermions. In this process, however,
another doubling of fermion modes is introduced in order to deal with
the positive definite square of the determinant instead of the
determinant itself.  The only known way to avoid this doubling is to
include the determinant in the observables and not use it as a
probability weight.  Thus, in the standard lattice formulation one
deals with an $n_f>1$ Schwinger model.  Recently some exact results
became available for the continuum mass\-less $n_f>1$ Schwinger model
\cite{GaSe94}, thus the situation in respect to the flavour problem of
lattice versions has somewhat improved.

The naive and the staggered formulation has a simple
phase diagram in the $(\be,m)$-plane, where  $\be=1/e^2$ is the gauge
coupling and $m$ denotes the bare fermion mass parameter.  The
continuum limit is recovered approaching the point $(\infty,0)$ along
trajectories with $m\simeq1/\sqrt{\be}$, where the proportionality
constant characterizes the physical mass of the bound state boson. In
that limit the known condensate value ${1\over e} \langle \psibar
\psi\rangle$ should be obtained.  Most of the lattice calculations
worked in that formulation and reproduced the expected continuum
results of the massless Schwinger model successfully (see e.g.
\cite{MaPaRe81,CaKe86,Di92}, for recent work on the non-compact
formulation cf. \cite{AzDiGa94}).

Little is known for the Wilson formulation, though. Here the two bare
coupling parameters are the gauge coupling and the hopping parameter
$\ka$ related to the bare fermion mass.  Chiral symmetry is explicitly
violated for finite $\be$ and most likely restored in spontaneously
broken form in the continuum limit.  This deprives us of a suitable
order parameter. Also, for the $n_f=1$ model, there is no massless
Goldstone boson. A continuum limit should be obtained in the approach
to $(\be=\infty, \ka=1/4)$ along trajectories following fixed values of
physical (dimensionless) quantities.  In $QCD_4$ there is a line
$\kac(\be)$ where the pion mass vanishes, corresponding to the
situation of a vanishing bare fermion mass (due to the Wilson term the
bare quark mass is renormalized additively). In the Schwinger model we
may expect a similar behaviour and in this case the massless continuum
Schwinger model is obtained following such a curve $\ka=\kac(\be)$.

Despite good knowledge of the properties of the theory in the
continuum, it is still a challenge to clarify the $n_f=1$ model phase
diagram on the lattice in the Wilson formulation.  One cannot use
$\langle \psibar \psi\rangle$ as an order parameter nor can we
determine the mass of the boson state in a direct simulation.  In this
paper we therefore try to contribute to this issue in an analytic (at
$\be=0$ and $\infty$) study and a computer calculation (at $\be=1$ and
5) of the partition function itself. With help of the equivalence of
the strong coupling model to the 7-vertex model \cite{Sa91,GaLaSa92} we
construct the hopping expansion up to $O(\ka^{128})$. At non-zero $\be$
we directly determine the fermion determinant in the numeric
integration over the gauge field background.  The scaling of the
Lee-Yang zeroes indicates that, somewhat contrary to an earlier finding
\cite{GaLaSa92}, there is indeed a phase transition line for all
positive values of the gauge coupling $\be$.  A subset of the numeric
results has been presented elsewhere \cite{GaLa94a}.

\section{Action, partition function and its zeroes}

The action for the massive lattice Schwinger model in the Wilson
representation is given by
\begin{eqnarray} \label{action}
S(\ka,\be)&=&S_F(\ka) + \be S_G \; , \\
S_F(\ka) &=& \sum_{x \in \Lambda} \left( \ka \sum_\mu
(\bar \psi(x+\hat \mu)(1+\gamma_\mu) U^{\dagger}_{\mu} (x) \psi(x)
\right. \nonumber \\
& &\left. + \bar \psi (x) (1-\gamma_\mu)
U_{\mu}(x) \psi(x+ \hat \mu) ) - \bar \psi (x) \psi (x) \right)
\nonumber  \\
\label{fermaction}
&\equiv & \sum_{x,x^\prime \in \Lambda}\bar \psi (x)
[M_\Lambda(\ka,U)]_{x,x^\prime}\psi (x^\prime)  \; .
\end{eqnarray}
For the gauge fields $S_G$ is the standard Wilson action with the
lattice gauge coupling $\be$ and $M_{\Lambda}(\ka,U)$ denotes the Dirac
operator on the lattice $\Lambda$.  Eq.s (\ref{action}) and
(\ref{fermaction}) are formal expressions in the partition function of
the model, which, after Grassmann `integration', is given by
\begin{eqnarray}\label{partitionfcn1}
Z_{\Lambda}(\ka,\be) &=&
\int  d\mu (U) \det M_{\Lambda}(\ka,U) e^{-\be S_G(U)}\\
\label{partitionfcn2}
&\equiv&Z_{G,\Lambda}(\be)  \langle\;
\det M_{\Lambda}(\ka,U) \rangle_G\; .
\end{eqnarray}
$Z_{G,\Lambda}(\be)$ is the partition function of the purely bosonic
system and $\langle {\cal O}(U)\rangle_G$ the expectation value of some
operator for that system. Without loss of generality we
normalize $Z_{G,\Lambda}(0)=1$.
The partition function $Z_{\Lambda}(\ka)$ is
a polynomial of degree $2 |\Lambda|$ in $\ka$ and thus an entire
function. All coefficients of that polynomial are positive. We know
that $\det M_{\Lambda}$ is strictly positive for
$\ka(\be)<\kac(\be=\infty)=1/4$; above this value of $\ka$ we
checked explicitly that individual configurations do produce negative
values although the partition function remains positive for finite
lattices.

% positivity of coefficients: e.g. from 7-vertex model
% positivity for k<1/4: Wilson formulation cannot have negative
% or zero eigenvalue for k<1/2d

The zeroes of that polynomial, called Lee-Yang zeroes
\cite{YaLe52}, have a non vanishing imaginary part for all
finite $|\Lambda|$. In the thermodynamic limit these zeroes pinch the
real axis of the complex $\ka$ plane and define thereby the point of
non-analyticity at $\kac$.  How the zeroes approach the real axis is
governed by finite size scaling, at a critical point related to the
critical exponents of the system \cite{ItPeZu83}.

{}From (\ref{partitionfcn2}) we see that the partition function of the
full model is proportional to the purely bosonic expectation value of
the determinant of the lattice Dirac operator. Since
$Z_{G,\Lambda}(\be) > 0$ we find that $Z_{\Lambda}(\ka,\be)$ is zero
only where $\langle \det M_\Lambda(\ka,U) \rangle_G = 0$.  Whereas mass
gap calculations cannot be performed for one flavour, this operator can
be calculated and this calculation can be extended to any number of
flavours.

For given $\be$ we have to determine the expectation value for the
determinant for complex $\ka \in \myc$. Only at $\be=0$ and
$\be=\infty$ this may be done analytically. Elsewhere one may rely on
the following numerical methods.
\begin{description}
\item[analytic continuation: ] One obtains values of $\langle \det
M_\Lambda \rangle$ for various real $\ka$. Performing a polynomial fit
and using this fit one analytically continues to complex $ \ka$.  In a
test for the controllable situation of free fermions we find unstable
results; even the closest zeroes could not be identified reliably.  A
better alternative is to obtain the coefficients of the polynomial
expressed in terms of moments by direct numerical simulation.  This
approach has been successfully used in a recent study in 4D
\cite{Barbour92}. The crucial point is the convergence of the
coefficients with $n$. This approach is closest related to the quite
successful method of analytic extrapolation and determination of zeroes
via histograms \cite{KeLa93}.
\item[direct evaluation: ] One integrates over the gauge fields with
the usual Monte Carlo simulation and produces gauge configurations with
the standard bosonic measure $e^{-\be S_G(U)}d\mu (U)$.  For each
configuration one determines $\det M_\Lambda $ for a sufficiently dense
set of complex $\ka$ in the expected region of the closest zero.  The
gauge field average eventually produces the required results.  This
method is applicable for small lattices only, since one has to
determine the determinant for each gauge field configuration
explicitly. On the other hand it does not involve any extrapolation and
works for arbitrary number of flavours.
\end{description}

\section{Analytic results}

\subsection{Strong coupling limit ($\be=0$)}

For small lattice size (e.g. $2\times 2$) one may analytically solve
\begin{equation}
\det M_{\Lambda}(\ka,U) = \sum_{n=0}^{2|\Lambda|} c_n(U) \ka^n
\end{equation}
and integrate over the gauge fields explicitly to obtain
\begin{equation}
Z_\Lambda(\ka)=\langle \det M_{\Lambda}(\ka,U) \rangle_G
\end{equation}
as a polynomial in $\ka$. Although done with symbolic programs
this becomes prohibitive for larger lattices.

Another method is based on a map of the massive lattice Schwinger model
(at $\be=0$) on the eight-vertex model \cite{Sa91}. Integrating over
the gauge fields first one finds a theory of non-intersecting loops
represented by a vertex model,
\begin{equation}
Z_{8V,\Lambda}(M) =
\sum_{ \{n_i\} } \prod^8_{i=1} \alpha_i^{n_i}\; ,
\end{equation}
where $\alpha_i$ denotes the coupling for vertex type $i$ and $n_i$ the
multiplicity. The Schwinger model corresponds to $\alpha_1=M^2$,
$\alpha_2=0$, $\alpha_3=a_4=1$, $\alpha_{i>4}=\ot$ and $M=1/(2\ka)$.
In particular the vertextype 2 corresponding to the intersection of
loops is forbidden. We then produce all possible configurations summing
up the corresponding weights.

A configuration may be represented by a legal set of vertices or,
equivalently, by a collection of non-intersecting closed loops (out of
connected links).  For the determination of the series we
simultaneously use the link- and the vertex representation. For
lattices of size $L\times N$ (periodic b.c.) the number of a priori
possible link configurations is $2^{2LN}$.

The transfer matrix approach is  by far the most economic approach.
For a lattice of size $L\times N$ we consider the transfer matrix for a
column of $L$ vertices, $T(a,b)$ where $a, b$ denote the link
configurations of the left and righthand set of $L$ links each, e.g.
$a \equiv (a_1,a_2, \ldots ,a_L)$; $a$ and $b$ may assume $2^L$ values
each.  The internal variables have been integrated and periodic b.c.
conditions in the vertical direction have been implemented. Each entry
in $T$ is given by a polynomial term $M^{2n_1}$ if it has $n_1$
vertices of type 1.

One has the symmetries $T(a,b)=T(b,a)$
$=T(\bar{a},\bar{b})=T(\bar{b},\bar{a})$, with the notation $\bar a
\equiv (a_L, \ldots ,a_2, a_1)$, reversal of the direction.  Furthermore
only entries with even number of occupied external links (corresponding
to values of $a$ and $b$) and allowed vertex configurations are
non-vanishing. This effectively reduces the number of relevant entries
by a factor of roughly 8. In the computer implementation only relevant
entries of $T$ are kept. For $L=8$, $T$ has only 8384 independent
non-zero entries, each a polynomial in $M$.  From $T$ one constructs
\begin{equation}
T^2(a,c) = \sum_b T(a,b) T(b,c) \quad , \quad
 T^4 = T^2.T^2\, ,
\end{equation}
and so on. Then
\begin{equation}
Z_{8V}(M;L\times N)= \mbox{Tr}\{ T^N\} = \sum_a T^N(a,a)\; .
\end{equation}

Alternatively, we also considered blocks of $4\times 4$, which may be
described by functions, $P(a,b,c,d)$, with four arguments denoting the
link configurations of the four edges. Again one has various rotational
and reflection  symmetries reducing the number of independent non-zero
entries of $P$, which are constructed explicitly at the begin. From
$P(a,b,c,d)$ one may construct the series for e.g. $4\times 8$ and
$8\times 8$ lattices by suitable summations, like
\begin{equation}
Z_{8V}(M;8\times 8)=\sum_{abcdrstu}
P(a,b,c,d) P(a,u,c,t) P(r,b,s,d) P(r,u,s,t)\; .
\end{equation}
This method is less efficient than the one-column transfer matrix
approach, which we finally used to construct the series.

In the 7-vertex model the partition function is given as a series in
the variable $M$, $Z_{8V}(M; L\times N)=\sum_n d_n M^n$.  In the 2D
Schwinger model one uses the hopping parameter $\ka$ and in table
\ref{seriescoeffs} we give the coefficients of the series in this
variable,
\begin{equation}
Z_{\Lambda}(\ka) = ({2\ka})^{LN} Z_{8V}({1\over 2\ka}
; L\times N) = \sum_n c_n \ka^n\; ,
\end{equation}
for various lattice
sizes.  The coefficients $c_n$ are integers and are related to the
coefficient of the series in $M$ through $c_n = 2^n d_{LN-n}$.  Fig. 1a
give the positions of the zeroes in the first quadrant of $\myc$ for
lattice size $8\times 8$.

\subsection{Free fermions ($\be=\infty$)}

For $U(1)$ gauge systems with torus geometry the choice between
periodic or antiperiodic boundary conditions (b.c.) for the lattice
Dirac operator $M_\Lambda$ is irrelevant for all beta. Assume a field
configuration in fixed gauge, then antiperiodic b.c. amount only to
multiplying the gauge fields at the boundary with a factor of $-1 \in
U(1)$.  This new configuration is within the sum over all gauge field
configurations with the same weight as the original one.  Thus it is a
symmetry of the gauge field integral.

At  $\be=\infty$, if we fix the gauge, the choice of b.c. may be
parametrized by two $U(1)$ group elements $(A,B)$, e.g. for
antiperiodic b.c.  $A=B=-1$. Thus, as is done usually in analytic
calculations, we may discuss the system of {\em free} fermions with
antiperiodic b.c. as one particular configuration of the $\be=\infty$
limit. In this case the fermionic action can be diagonalized and the
determinant evaluated by Fourier transformation. Within the limitations
of available workstations this may be done with symbolic computer
programs up to lattice sizes $32 \times 32$ and larger. However, the
results for the complex positions of zeroes are quite different for
periodic and antiperiodic b.c. although the scaling behaviour agrees.
Due to the dependence on the b.c. we cannot compare the zeroes directly
with our results at finite $\be$.

Unlike $\be<\infty$ in the free case the zeroes occur degenerate with
multiplicity 4 and 8; fig. 1b shows the zeroes (in the first quadrant
of $\myc$) at $\be=\infty$ for lattice size $8\times 8$.  For both, at
$\be=0$ and $\be=\infty$, we find a distribution of zeroes, that does
not follow a simple geometric shape. The general shape of the
distribution at $\be=\infty$ becomes clearer at larger lattices, as
shown in fig. 1c for lattice size $32\times 32$.

\section{Numerical results and discussion}

For $0<\be<\infty$ we use numeric techniques to obtain information on
the closest zero on lattice of size $2\times 2$, $4\times 4$, and
$8\times 8$, following the second method discussed at the end of
sec.2.  We simulated background gauge field configurations for $\be =
5, \; \be = 1$, and also for $\be = 0$, in order to check the
reliability of the approach.  For each configuration we evaluated the
determinant on a grid of $20\times 20$ complex $\ka$ values in the
presumed region of the closest zeroes. The final expectation values on
this grid are then analyzed with help of interpolation.  For lattice
$2\times 2$ size we summed over 10000 gauge field configurations, for
$4\times 4$ and $8\times8$ over 5000.

The analytic results at $\be=0$ are in excellent agreement with the
numeric results for lattice sizes $2\times 2$ and $4\times 4$. At the
largest lattice size, however, the exact position is off from the
numerically determined position by 2 standard deviations (the jackknife
statistical error).  Actually, $\be=0$ is the worst case in the gauge
field integration, since the configurations are completely random.  We
have to conclude that for the largest lattice size the number of gauge
configurations considered is too small. (A test at $\be=0$ with a
statistics of 32000  did not sufficiently improve the situation.) For
this reason for lattice size $8\times 8$ we doubled the statistical
errors obtained at $\be=0$ to define the errorbars for the other
$\be$-values and consider these values with some caution.

Fig. 2 summarizes our results for the size- and $\be$-dependence of the
zeroes closest to the real $\ka$-axis.  The imaginary parts show the
tendency to vanish for $L\to \infty$ indicating the existence of a
phase transition for all $\be$ along a curve $\kac(\be)$.  For free
fermions, where $\ka-\kac\simeq1/\xi$, one finds $O(1/L)$ dependence.
For $\be<\infty$, for the lattices considered, the behaviour indicates
an even faster approach towards zero, like the $O(1/L^2)$ expected at
first order transitions (where the susceptibility $\simeq L^2$).
However, the errorbars and smallness of the lattices do not justify
stronger statements.  Also, considering the non-uniform behaviour of
the real parts shows that we are not yet in the asymptotic regime.

As mentioned, we know of no suitable order parameter to identify
existence and type of the possible phase transition line.  Earlier
investigations of the behaviour of the strongly coupled massive lattice
Schwinger model based on the 7-vertex model \cite{GaLaSa92} rather
suggested that there is no second critical point at $\be = 0$ for $\ka
< \infty$. To clarify this inconsistency we repeated that 7-vertex
calculation with significantly increased statistics (up to a factor of
ten) and denser grid in the coupling. It turned out that with this
improvement one does indeed find indications of scaling in the
susceptibility for lattices larger then $|\Lambda| = 32 \times 32$.
There is still a controversy concerning the peak position and the
boundary conditions of the 7-vertex model to be settled.

In summary, the $n_f=1$ massive (and massless) Schwinger model on the
lattice and with Wilson fermions is still a formidable task. Direct
simulation of the model and determination of the bosonic mass seem to
be impossible with present day resources.  From the equivalence to the
8-vertex model we know that there is an isolated critical point at $\be
= 0$ and $\ka = \infty$ \cite{Sa91,GaLaSa92}; this is most likely not
the endpoint of the singular line discussed above.  The expansion in
$\ka$ around $(\be =0, \ka =0)$ is convergent only for $|\ka| <
\bar \ka \leq 1/2$\cite{GaLaSa92}.  The responsible singularity could
lie at any complex $\ka$ with $|\ka| = \bar \ka$ and therefore  $\bar
\ka$ does not necessarily define a critical point.  The model is
believed to have a line of phase transitions at $\ka_c(\be)$ beginning
at $\ka_c(\infty) = 1/(2d)$ which is a point with a second order
transition and running to some $\ka_c(0) < \infty$. Except for $\be
\rightarrow \infty$, the free model, there is no proof that this
statement is correct.

Our findings from the direct evaluation of the partition function and
the finite size dependence of the Lee-Yang zeroes support the scenario
of a phase transition line from $\be=0, \ka\simeq 0.38(2) $ up to
$\be=\infty,\ka=1/4$.  The phase diagram may look similar to the
$n_f>1$ situation, also for QCD$_4$.

\section*{Acknowledgement}
We wish to thank C. Gattringer, R. Kenna, M. Salmhofer and E. Seiler
for many fruitful discussions.

%\bibliographystyle{npb}
%\bibliography{/usr/people/cbl/tex/refs/lgt}
%\end{thebibliography}

\newpage
\begin{table}
\caption{Series for $Z_\Lambda(\ka; L\times N) =
\sum_n c_n \ka^n $.\label{seriescoeffs}}
\begin{center}
\begin{tabular}{rrrrrr}
\hline
n& $2\times2$ &$4\times4$& $8\times8$\cr
0 &1  &  1 &  1 \cr
4 &64 &&\cr
 8 &768 & 2304 & 1024 \cr
 12 && 32768 & 32768 \cr
 16 &&  2617344 & 2547712\cr
 20 && 51904512  & 121634816\cr
 24 && 1068498944  & 6214909952 \cr
 28 && 9663676416  & 291286024192\cr
 32 &&  37597741056  & 16264921612288 \cr
 36 && & 733659871051776 \cr
 40 &&&  32983549241982976 \cr
 44 &&& 1393668629299462144\cr
 48 &&&  57045024334275411968\cr
 52 &&&   2178933957104869834752 \cr
 56 &&&  78035756256081760747520 \cr
60 &&& 2591089287867411030081536\cr
64 &&& 79973512017556234532028416\cr
68 &&  &     2269245313085843057330356224  \cr
72 &&  &    58685404081509064264407580672  \cr
76 &&  &    1378038126803683051771720105984  \cr
80 &&  &    29110687425223635390841502040064  \cr
84  && &     547924220534282839622026012917760  \cr
88  && &     9113040973297576534851687523811328  \cr
92  && &    132393136404532724219119019631837184  \cr
96  && &    1658720256337887142482109949475291136  \cr
100  && &    17660710533168127842880532530400854016  \cr
104  &&  &     157058824477979862076814317113632096256  \cr
108  &&  &     1140500578260786292933495730676818771968  \cr
112  &&  &     6579687246247908061970155944177873977344  \cr
116  &&  &     28956384933789738373830226402627572203520  \cr
120  &&  &     92150412611536628441916337226044921610240  \cr
124  &&  &    187529160722546587292381268726888701886464  \cr
128  &&  &     223167089080216837357579582404487997816832 \cr
\hline
\end{tabular}
\end{center}
\end{table}
\clearpage
\newpage

\section*{Figures}

{\noindent \bf Fig 1:}
Distribution of the complex zeroes for (a) $\be=0$ on an $8\times 8$
lattice and at $\be=\infty$ on (b) $8\times 8$ and (c) $32\times 32$
lattices.\\
{}~\\
{\noindent \bf Fig 2:}
(a) The zeroes closest to the real $\ka$-axis at $\be=0$ (squares),
$\be=1$ (crosses) and  $\be=5$ (triangles); subsequent lattice sizes
($2\times2$, $4\times 4$, $8\times 8$) are connected by lines to guide
the eye.  (b) Scaling behaviour of the imaginary parts of the closest
Lee-Yang zeroes (notation as in (a)) vs. $1/L$; we also show results
for $\be=\infty$ and antiperiodic b.c. as discussed in the text.

\end{document}